\documentclass[showpacs]{revtex4}
\usepackage{graphicx}

\begin{document}
\title{Binding Effects in Multivalent Gibbs-Donnan Equilibrium}
\author{M. Castelnovo}\altaffiliation{Laboratoire Joliot-Curie et Laboratoire de Physique -- Ecole Normale Sup\'erieure de Lyon, 46 All\'ee d'Italie, 69364 LYON CEDEX 07, FRANCE} 
\author{ A. Evilevitch}
\altaffiliation{Department of Biochemistry, Center of Chemistry and Chemical Engineering, Lund University, P.O. Box 124, S-221 00 LUND, SWEDEN }
\begin{abstract}
The classical Gibbs-Donnan equilibrium describes excess osmotic pressure associated with confined colloidal charges embedded in an electrolyte solution. In this work, we extend this approach to describe the influence of multivalent ion binding on the equilibrium force acting on a charged rod translocating between two compartments, thereby mimicking ionic effects on force balance during \textit{in vitro} DNA ejection from bacteriophage. The subtle interplay between Gibbs-Donnan equilibrium and adsorption equilibrium leads to a non-monotonic variation of the ejection force as multivalent salt concentration is increased, in qualitative agreement with experimental observations.
\end{abstract}

\maketitle

\section{Introduction}
The classical Gibbs-Donnan equilibrium is usually invoked to explain various behaviors in ionic systems: swelling of polyelectrolyte gels, membrane potentials  and osmotic shock experiments of cells or viruses \cite{refgen}. The physics behind these phenomena is associated with some colloidal charge (usually synthetic or bio-polyelectrolytes) confined within a subpart of the system, while small co- \textit{and} counter-ions can diffuse freely in and out of this region \cite{donnan}. Due to the electroneutrality inside the compartment, this confinement leads to some excess of counterions relative to the outer compartment, and therefore an additional osmotic pressure is set up inside. This effect is nowadays well-known and characterized experimentally on model systems \cite{raspaud}. 

From a theoretical point of view, the Gibbs-Donnan approach is a simple and versatile way of incorporating electrostatic effects up to leading order 
in \textit{inhomogeneous} systems, through the use of ionic chemical potential balance between two \textit{homogeneous} phases and the electroneutrality condition. In systems either where colloidal charge is large or where the buffer solution contains multivalent ions, the Gibbs-Donnan theory has to be extended to take into account the electrostatic binding of counterions onto the colloidal charge \cite{reiss}. Motivated by recent \textit{in vitro} experiments measuring multivalent salt influence on the force ejecting DNA from bacteriophage Lambda \cite{alexnew}, we propose in this Letter such an extension in order to include multivalent ion binding effects in the classical Gibbs-Donnan approach. Indeed, we expect the interplay between Z-valent ion partitioning between inside and outside the viral capsid (pure Donnan effect), and the binding statistics onto DNA inside and outside (pure Langmuir-like adsorption statistics), to give raise to non-trivial translocation behaviors. In a slightly different context for example, the balance of these two effects has been shown by Klein Wolterink \textit{et al.} to produce non-monotonic variation of the radius of gyration of a strongly charged polyelectrolyte star with respect to salt concentration \cite{borisov}. As will be shown in this work, this simple generalization of Gibbs-Donnan approach is able to explain at least qualitatively the non-monotonicity of ejecting force as function of multivalent salt concentration. In particular, we find that the repulsive force decreases and then increases, upon increase in the added salt concentration. The minimum in the force is associated to the charge neutralization point of DNA, which correlates well with experimental results \cite{alexnew}. 

\section{Model}
\begin{figure}
\includegraphics[scale=0.9]{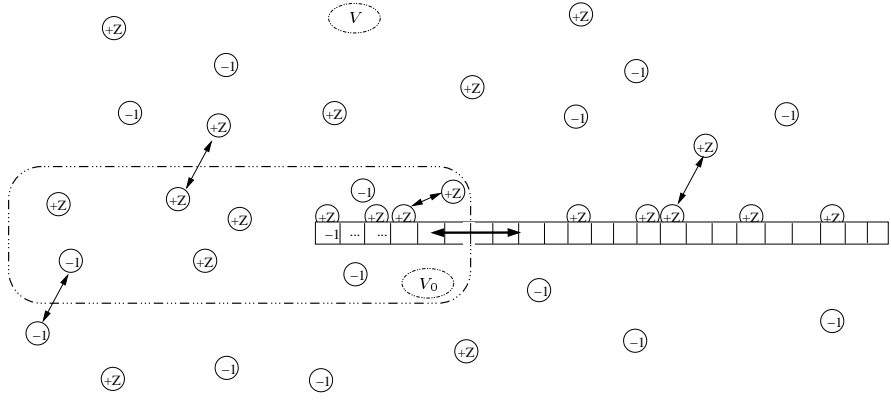}
\caption{Sketch of the model. The two-sided arrows indicate different equilibria taking place in the system}
\label{cartoon}
\end{figure}
The model system considered in this work is depicted in figure \ref{cartoon}. It consists of a cavity of volume $V_0$ embedded into a larger volume $V>>V_0$. The whole volume $V$ is filled with a (Z:1) electrolyte at an adjustable concentration. The cavity is permeable to both species (Z-valent cations and monovalent anions) of the electrolyte. A translocation gate on the cavity allows a rigid rod bearing $L$ uniformly distributed negative charges to move between inside and outside the cavity. For a given configuration, there are $L_{in}$ and $L-L_{in}$ negative charges respectively inside and outside the cavity. Multivalent cations are expected to bind onto the charged rod, due to strong electrostatic interactions. The free energy of this system is written as $F_{tot}=F_{layers}+F_{free}+F_{neutral}$ where the first term is associated with the  Z-cation adsorbed layers on the rod, the second term is associated to remaining free ions in the solution, and finally the last term takes into account any non-ionic effects. It can include for example the effect of a neutral osmotic pressure difference between inside and outside the cavity due to neutral polymers that cannot enter the cavity, as in the experimental setup used by Evilevitch \textit{et al.} \cite{alexold,alexnew}, or bending effects in the case where the rod has a finite bending modulus \cite{rob1,rob2,tzlil}. 

Following the spirit of the original Gibbs-Donnan approach, the specific features of electrostatic binding are neglected, and the binding of Z-valent cations onto the rod is approximated by Langmuir-like adsorption isotherms \cite{donnan}. Each negative charge of the rod is therefore assumed to be a potential discrete binding site for a single Z-valent cation. The energy gain associated with a single binding event is given by $-kT\epsilon_A$. This energy is of order $\epsilon_A\simeq Zl_B/b$, where $b$ is the typical size of rod unit and $l_B$ the Bjerrum length. This neglects any correlation effects among adsorbed Z-valent ions. It can be checked \textit{a posteriori} that including a certain level of correlations does not change the qualitative picture drawn by this simple model. Denoting by $N_{Ain}$ and $N_{Aout}$ respectively the number of adsorbed ions on the rod inside and outside the cavity, the free energy of the adsorbed layers is $F_{layers}=F_A(N_{Ain},L_{in})+F_A(N_{Aout},L-L_{in})$, where the adsorption free energy of ions on a one dimensional lattice reads 
\begin{equation}
\frac{F_A(N_A,L)}{kT}=N_A \ln \frac{N_A}{L}+ (L-N_A) \ln \left(1-\frac{N_A}{L}\right)-N_A\epsilon_A
\end{equation} 
The first two terms describe the mixing entropy of adsorbed ions, and the last one is the adsorption energy. 
Assuming a unique value $b^3$ for the molar volume of each species (Z-cations, anions, and negative charges on the rod) for the sake of simplicity, the free energy of remaining free ions in the solution is given by the sum of translational free energies $F_T(N,V)=kTN(\ln (Nb^3/V)-1)$ of Z-cations and anions, respectively, inside and outside the cavity: 
\begin{eqnarray}
F_{free} & = & \bigg\{ F_T(N_{Zin}-N_{Ain},V_0-b^3L_{in})+F_T(ZN_{Zin}-L_{in},V_0-b^3L_{in})\bigg\}\nonumber\\
 & & +\bigg\{F_T(N_Z-N_{Zin}-N_{Aout},V-V_0-b^3(L-L_{in}))\nonumber \\
& & +F_T(Z(N_Z-N_{Zin})-(L-L_{in}),V-V_0-b^3(L-L_{in}))\bigg\}
\end{eqnarray}
Note that the number of Z-cations inside the cavity is $N_{Zin}$, including both free and adsorbed ions, while the total number of Z-cations in the solution (both inside and outside the cavity, free and adsorbed) is $N_Z$. Following Donnan, electroneutrality conditions have been taken into account inside and outside the cavity, so that the numbers of inner and outer anions are $N_{-in}=ZN_{Zin}-L_{in}$ and $N_{-out}=Z(N_Z-N_{Zin})-(L-L_{in})$. 

At equilibrium, the free energy of the system is minimum with respect to $N_{Ain},N_{Aout},N_{Zin}$ and $L_{in}$. The first three minimization conditions mean simply that the adsorbed layers are in equilibrium with the free Z-ions of the solution, both inside and outside the cavity (Langmuir balance), and that free Z-ions inside the cavity are in equilibrium with free Z-ions outside the cavity (Gibbs-Donnan balance). These equations are written in the limit $V>>V_0$, introducing species concentration $n_{Zin}=N_{Zin}/V_0,n_Z=N_Z/V,\phi_{in}=N_{Ain}/L_{in},\phi_{out}=N_{Aout}/(L-L_{in}),l_{in}=L_{in}/V_0,l=L/V$
\begin{eqnarray}
\label{phiin}\frac{\phi_{in}}{1-\phi_{in}} & = & (n_{Zin}-\phi_{in}l_{in})b^3e^{\epsilon_A}\\
\label{phiout}\frac{\phi_{out}}{1-\phi_{out}} & = & (n_{Z}-\phi_{out}l)b^3e^{\epsilon_A}\\
\label{donnan1}\left(n_{Zin}-\phi_{in}l_{in}\right)\left(Zn_{Zin}-l_{in}\right)^Z & = & \left(n_Z-\phi_{out}l\right)\left(Zn_{Z}-l\right)^Z
\end{eqnarray}
For a given value of $l_{in}$, these equations are solved self-consistently and their solutions allow calculation of properties of interest for the translocation problem, like the ionic contribution to the force acting on the rod $f_{ionic}=\frac{\partial (F_{layers}+F_{free})}{\partial L_{in}b}$.
Under equilibrium conditions, the value of $l_{in}$ is set by minimization of the free energy with respect to $L_{in}$. Introducing the force contribution associated with non-ionic features $f_{neutral}=\frac{\partial F_{neutral}}{\partial L_{in}b}$, this equation is simply the force balance on the rod $f_{neutral}+f_{ionic}=0$. In this Letter, we focus on the behavior of the ionic force for different solution conditions at fixed $l_{in}$, rather than determining the equilibrium partition of the rod inside and outside the capsid. Indeed, the value of $l_{in}$ is dependent on the choice of neutral force contribution $f_{neutral}$, which is beyond the scope of this work \cite{alexnew}.

\section{Results and discussion}
\begin{figure}
\includegraphics[scale=0.8]{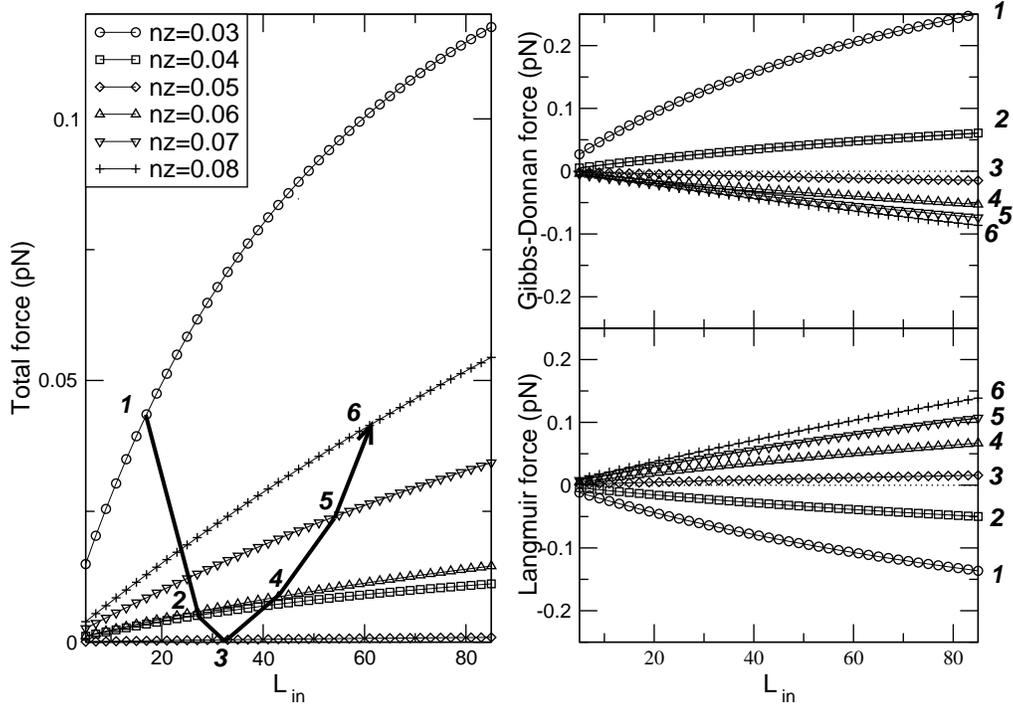}
\caption{Ionic force acting on the rod \textit{vs} rod length inside $L_{in}$, for different multivalent salt concentration. Parameters: $L=100,V_0=10^3b^3,V=10^9b^3,b=2nm,Z=4,\epsilon_A=2.5$. \textit{Left panel}: Total force; the large arrow and numbers highlight the variation of the force-length plots as multivalent salt concentration is increased. Note that the force decreases at first ($1\rightarrow 2 \rightarrow 3$) upon increase in $n_Z$, and then increases ($4\rightarrow 5 \rightarrow 6$). \textit{Upper right panel}: Gibbs-Donnan contribution to the ionic force. \textit{Lower right panel}: Langmuir contribution to the ionic force.}
\label{plotforce}
\end{figure}

\begin{figure}
\includegraphics[scale=0.8]{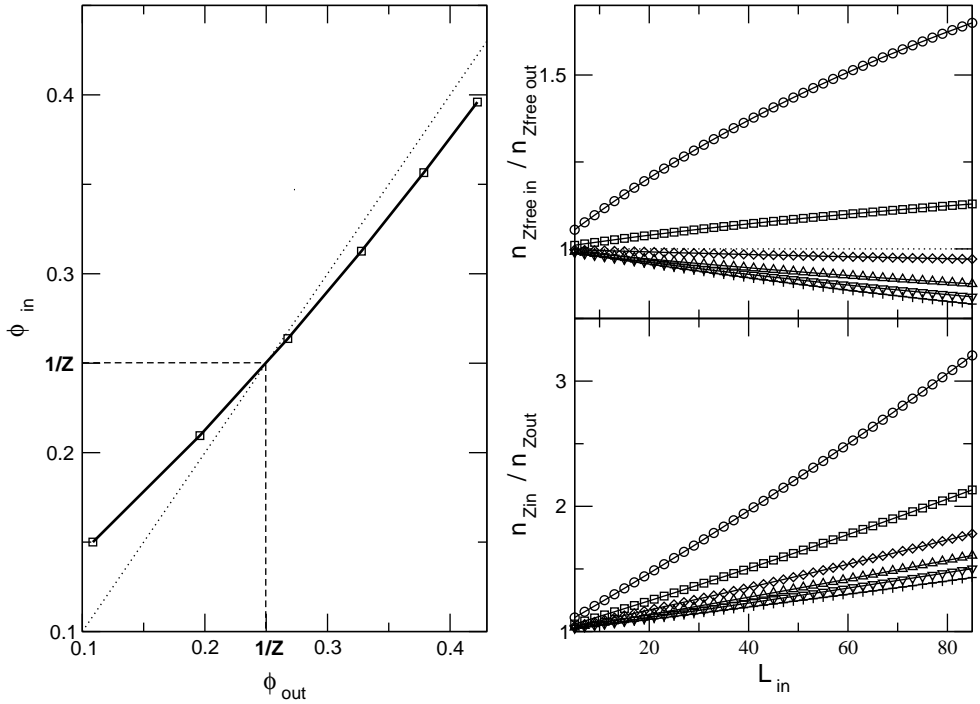}
\caption{Adsorbed and free Z-cation concentrations. The value of parameters are the same as in figure \ref{plotforce}. \textit{Left panel}: Inner \textit{vs} outer adsorbed degree of Z-cations for concentrations $n_Z b^3=0.01,0.02,0.03,0.04,0.05,0.06$ ($\phi_{in,out}$ increase with $n_Z$).\textit{Upper right panel}: Ratio of free Z-cation concentration inside/outside \textit{vs} rod length inside. Symbols are defined in figure \ref{plotforce}.\textit{Lower right panel}: Ratio of total Z-cation concentration inside/outside \textit{vs} rod length inside. Symbols are defined in figure \ref{plotforce}.}
\label{plotconc}
\end{figure}

In the absence of any analytical solutions for the equilibrium equations \ref{phiin}-\ref{donnan1}, we solved them numerically for a given representative set of parameters.
As it is seen in figure \ref{plotforce}, the ionic force is always positive and increasing with respect to the rod length inside the cavity. 
This has to be contrasted with the net attractive force found in the work of Zandi \textit{et al.} \cite{roya}, where the binding particles are only present inside the cavity and are not allowed to bind on the outer part of the rod. Now, increasing multivalent salt concentration leads to a non-monotonic variation of the force at \textit{fixed inner length} in the present model, which is the main result of this work: the force is first decreasing at low salt concentration, and then increasing at higher salt concentration. The multivalent salt threshold corresponds to the neutralization point of the rod $\phi_{in,out}\sim 1/Z$, as it is seen on figure \ref{plotconc}. This behavior is interpreted below by the interplay between Gibbs-Donnan and Langmuir contribution to the ionic force.

The ionic force is rewritten as the sum of three main contributions $f_{ionic}=f_{GD}+f_{L}+f_{\Delta \Pi}$. The three terms are respectively associated with the Gibbs-Donnan force contribution, the Langmuir force contribution and the osmotic force \cite{castelosmotic}. These forces read
\begin{eqnarray}
\label{fgd}\frac{f_{GD}b}{kT} & = & -\ln \frac{Zn_{Zin}-l_{in}}{Zn_Z-l} \\
\label{fl}\frac{f_{L}b}{kT} & = & \ln\frac{1-\phi_{in}}{1-\phi_{out}}\\
\label{fo}\frac{f_{\Delta \Pi}b}{kT} & = & b^3(\left(n_{Zin}-\phi_{in}l_{in}\right)+\left(Zn_{Zin}-l_{in}\right)- \left(n_Z-\phi_{out}l\right)-\left(Zn_{Z}-l\right))
\end{eqnarray}
The first contribution Eq. \ref{fgd} is associated with the free ions partitioning between inside and outside the cavity. In the absence of ion binding on the rod, there is a depletion of anions and an excess of Z-cations inside the cavity, due to the negative charge of the rod. Within classical Gibbs-Donnan approach, the net force is repulsive. The effect of ion binding on the rod at low multivalent ion concentration is mainly to reduce the net negative charge of the rod, without affecting the sign of the force. Above the neutralization threshold concentration $n_Z^*$, for which $\phi_{out}\sim 1/Z$, the net charge of the rod is positive, due to the binding of Z-cations. In this case, there is a depletion of \textit{free} Z-cations and an excess of anions inside the cavity. However, in contrast to the previous case, there is still an overall excess of Z-cations inside (fig. \ref{plotconc}), a growing number of them being involved in the binding on the rod leading to its overcharging. The important proportion of adsorbed Z cations, not contributing to translational entropy inside the cavity, is now favorable to the rod being inside the cavity. As a consequence, the Gibbs-Donnan force is attractive as shown in figure \ref{plotforce}. This interpretation does not take into account the free energy of adsorbed layers, however,  which compensates this negative Gibbs-Donnan force to give a net repulsive force. 

Indeed the second contribution, Eq. \ref{fl}, comes from the mixing entropy of adsorbed ions along the rod. The sign of the force is mainly given by the relative value of inner and outer adsorption degrees $\phi_{in}$ and $\phi_{out}$. At low multivalent concentration, Gibbs-Donnan equilibrium favors free Z-cations excess inside, thereby increasing inner adsorption relative to the outside environment. Therefore the Langmuir force tends to pull the rod inside. 
This is the quasistatic effect seen in simulations of reference \cite{roya} for particular conditions, though with a larger amplitude, due to the asymetrical binding on the rod.
Above the neutralization threshold, the situation is reversed, and outer adsorption is favored $\phi_{out}>\phi_{in}$ (fig. \ref{plotconc}). As a consequence, Langmuir force is now repulsive, favoring rod ejection from the cavity.

The last contribution Eq. \ref{fo} in the ionic force is the osmotic force, first introduced in \cite{castelosmotic}, and it is proportional to the osmotic pressure difference between inside and outside the cavity. It arises from the fact that, in order to insert the rod which has a finite volume inside the cavity, one has to perform a pressure work. Within the conditions chosen in this work, this osmotic force is one order of magnitude smaller than Gibbs-Donnan and Langmuir forces, and is therefore disregarded in the present analysis. 

For the chosen values of parameters (rod length, adsorption energy, etc...), the magnitude of the ionic force is of order of a few tenths of a piconewton, which is quite small at the molecular level. This comes mainly from the fact that the net entropic forces discussed previously arise from ionic concentration gradients between two compartments and two parts of the rod, unlike asymetrical situations where binding is limited to the inner part of the rod for example. Increasing the length and adsorption energy at fixed $V_0$ produces larger ionic force amplitudes. Added to the neutral force contribution, the ionic force can still lead to substantial changing of behaviors, as is discussed below.

\section{Applications and limitations} 
As already mentioned in the introduction of this Letter, the present model was originally designed in order to evaluate the leading order of multivalent ionic effects on \textit{in vitro} DNA ejection from bacteriophage \cite{alexnew}. Within these experiments, DNA ejection is triggered using solubilized specific receptor in a solution containing neutral polymers producing an osmotic force that is able to balance the DNA ejection force. Changing ionic conditions of the solution at constant polymer concentration allows us to address the influence of ionic environment on the ejection force. The present model is of course a crude simplification of the real system. However it is able to predict qualitatively the non-monotonicity of ejection force, as it is observed in the relevant experiments. Within this model, the minimum in the force is associated to the neutralization of charged rod $\phi \sim 1/Z$. Using Eq. \ref{phiout}, the Z-valent threshold concentration scales like $n_{Z}\sim e^{-\epsilon_A}$. This implies that this threshold concentration decreases with the valency of counterions, a fact that is observed experimentally for three different valencies: $Na$ ($Z=1$), $Mg$ ($Z=2$), spermine ($Z=4$). Quantitative comparison of predicted and measured threshold is however difficult since the experimental buffer contains more than one multivalent salt. Our model can easily be extended to include more salt types and their binding competition with different energies $\epsilon_{A\, i}$ ($i=1,2...$).
 This modified model leads qualitatively to the same results as the one derived for one multivalent salt (data not shown). The quantitative difference between the two models is the precise location of ionic force minimum or neutralization threshold. Since the main effect is already observed within the present simple one-salt model, we prefer to stick to this model, for the sake of clarity.
Note that Gibbs-Donnan balance has already been used by Odijk and co-workers in order to describe DNA packing in a model-bacteriophage \cite{odijk3}. But the presence of multivalent ions as well as their influence on DNA ejection has been disregarded.

The main effect that is not included explicitly within this model is the presence of correlations between bound ions, changing the effective adsorption energy \cite{shklovskii}. However, we don't expect this effect to change drastically the results derived within our model: indeed, the non-monotonicity of the ionic force is related within our model to the presence of neutralization of the rod and possible overcharging, and not to its precise description.
We checked with non-linear adsorption energy that prohibits the rod from being unrealistically too overcharged, that the qualitative behavior is unchanged. In particular, we observed that including only correlations and neglecting Gibbs-Donnan balance does not lead to non-monotonicity of ionic force, so that the experimentally observed minimum of the force can not be explained solely by invoking correlation effects. Therefore both basic ingredients of our model, Gibbs-Donnan balance and adsorption statistics, are necessary to observe the non-monotonicity of the force. Notice that for similar reasons we don't expect correlations of free ions, like the leading order Debye-Huckel correlations \cite{refgen}, to change qualitatively the results obtained in this Letter.

In order to interpret the results of DNA ejection experiments within this model, an additional implicit assumption has been made: the ionic effects can be decoupled from other contributions, termed ``neutral'' within this work. This assumption is valid within the simple geometry shown in figure \ref{cartoon}. In a more realistic model of viral bacteriophage, where DNA is arranged in a spool-like fashion within the viral cavity, the neutral contribution would include bending energy as well as direct electrostatic repulsion between neighbouring DNA turns in the spool \cite{rob1,rob2,tzlil}. Up to now except in aforementioned works of Odijk and co-workers  \cite{odijk3}, the inclusion of electrostatic effects in such models has been based on phenomenological expressions describing osmotic pressure of hexagonal phases of DNA as measured in the experiments of \cite{raspaud2} for example. Since the simple extended Gibbs-Donnan model presented here is qualitatively consistent with these expressions\footnote{M. Castelnovo, unpublished results}, we expect our model to describe correctly the leading order of multivalent ionic effects.

The range of application of the simple model proposed in this Letter is larger than the ejection experiment just mentioned.  Indeed it is relevant to the description of translocation properties of polyelectrolytes\cite{mutu}, whenever the separating interface is permeable to mobile free ions. We showed in this paper that the force balance acting on the rod can be displaced by changing ionic conditions. Within a biological context, the applicability of the present model for the translocation of bio-polyelectrolytes between different cell compartments, for example, is limited by the presence of ionic channels on the interface (cell compartment membrane) that actively regulate  the ionic gradients accross the interface. Therefore the pure Gibbs-Donnan balance is no longer obtained, and new models have to be developed.

\acknowledgments
Fruitful discussions at early stages of this work with W.M. Gelbart are greatfully acknowledged.

\end{document}